%% file: scenegcn.tex
\documentclass[journal]{IEEEtran}

\usepackage{times}
\usepackage{epsfig}
\usepackage{graphicx}
\usepackage{amsmath}
\usepackage{amssymb}
\usepackage{subfigure}
\usepackage{color}
\usepackage{enumerate}
\usepackage{booktabs,multirow,tabularx}
\usepackage[bookmarks=true]{hyperref}

\usepackage[utf8]{inputenc} % allow utf-8 input
\usepackage{url}            % simple URL typesetting
\usepackage{booktabs}       % professional-quality tables
\usepackage{amsfonts}       % blackboard math symbols
\usepackage{nicefrac}       % compact symbols for 1/2, etc.
\usepackage{microtype}      % microtypography
\usepackage{threeparttable}
\usepackage{color, colortbl}

\definecolor{lightblue}{rgb}{0.93,0.96,0.9765}
\definecolor{lightyellow}{rgb}{1,1,0.88}
\newcommand{\etal}{$et$ $al.$}

\title{Scene Graph Reasoning with Prior Visual Relationship for Visual Question Answering}

\author{
Zhuoqian~Yang,
Zengchang~Qin,
Jing~Yu,
Yue~Hu% <-this % stops a space
\thanks{Zhuoqian Yang and Zengchang Qin are with Intelligent Computing and Machine Learning Lab, School of Automation Science and Electrical Engineering, Beihang University.}
\thanks{Jing Yu and Yue Hu are with the Institute of Information Engineering, Chinese Academy of Sciences.}
}

\begin{document}

\maketitle

\begin{abstract}
One of the key issues of Visual Question Answering (VQA) is to reason with semantic clues in the visual content under the guidance of the question, how to model relational semantics still remains as a great challenge. To fully capture visual semantics, we propose to reason over a structured visual representation -- scene graph, with embedded objects and inter-object relationships. This shows great benefit over vanilla vector representations and implicit visual relationship learning.
Based on existing visual relationship models, we  propose a visual relationship encoder that projects visual relationships into a learned deep semantic space constrained by visual context and language priors. Upon the constructed graph, we propose a \textbf{Scene} \textbf{G}raph \textbf{C}onvolutional \textbf{N}etwork (\textbf{SceneGCN}) to jointly reason the object properties and relational semantics for the correct answer. We demonstrate the model's effectiveness and interpretability on the challenging \textit{GQA} dataset and the classical \textit{VQA 2.0} dataset, remarkably achieving state-of-the-art 54.56\% accuracy on GQA compared to the existing best model. 
\end{abstract}


\input{sections/1_introduction}
\input{sections/2_related_works}
\input{sections/4_methodology}
\input{sections/5_experiments}
\input{sections/6_conclusion}

\bibliographystyle{unsrt}
\bibliography{ref}

% \appendix
% \include{sections/7_supplementary} 

\end{document}

%% file: sections/1_introduction.tex
\section{Introduction}

\begin{figure*}[!t]
\centering
\includegraphics[width=0.9\textwidth]{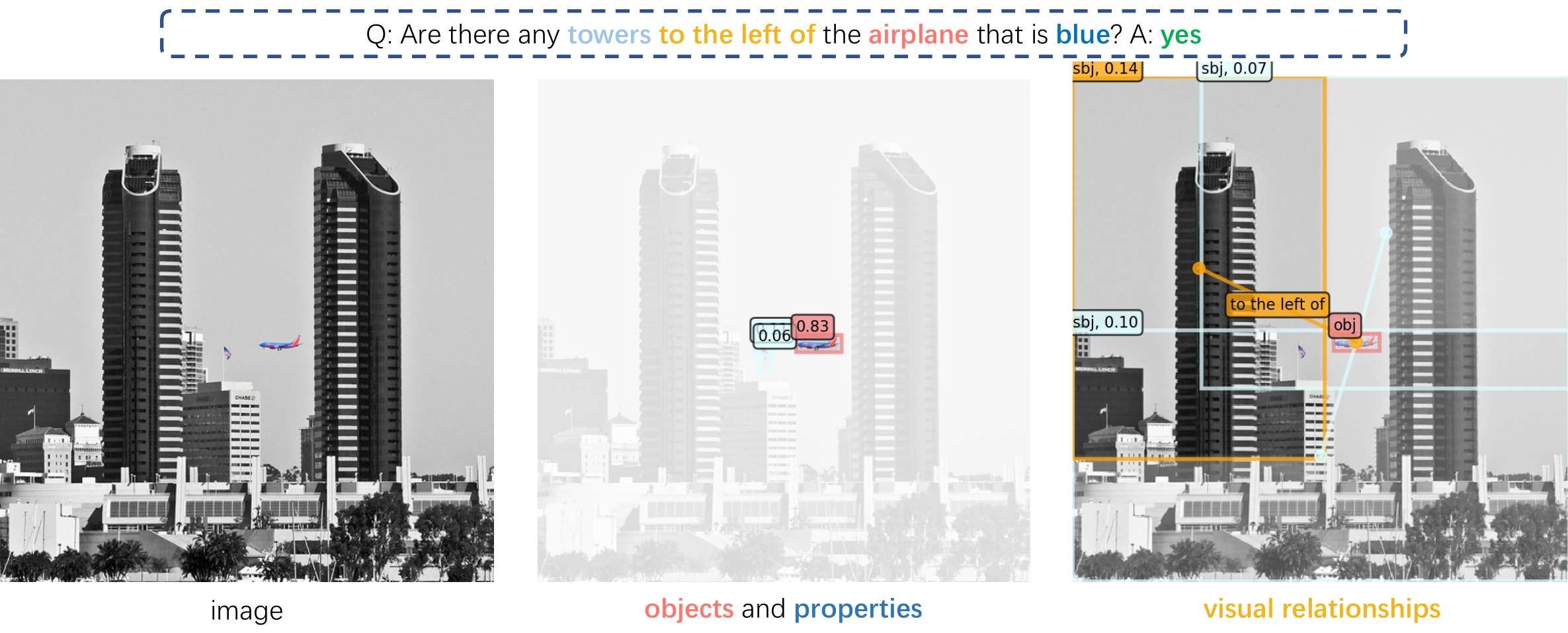} 
\caption{Semantic visual relationships are crucial in inferring the correct answer. The question requires not only the recognition of the object \textit{plane} that is \textit{blue}, but also the relationship $\left\langle\textit{tower}, \textit{to the left of}, \textit{plane}\right\rangle$. The proposed Scene Graph Convolutional Network (SceneGCN) is capable of locating the visual relationship that is closely related to the question with interpretable attention-based rationale.}
\label{fig:motivation}
\end{figure*}

Visual Question Answering (VQA) is one of the most challenging tasks in cross-modal information modeling: an image and a free-form question in natural language are presented to an intelligent agent who is required to determine the correct answer using both visual and textual information. 
A key issue in visual question answering is to reason with the semantic clues from the visual content under the guidance of the question. To achieve such human-like abilities, the intelligent agent should not only infer clues from individual objects but also the visual relationships between them. For example, to solve the VQA problem in Figure \ref{fig:motivation}, intelligent agents need to capture not only the object \textit{plane} that is \textit{blue}, but also the relationship $\left\langle\textit{tower}, \textit{to the left of}, \textit{plane}\right\rangle$ for correct answer prediction.

% \begin{figure}
% \centering
% \includegraphics[width=0.5\textwidth]{figures/motivation.jpg} 
% \caption{Semantical visual relationships are crucially important in inferring the correct answer. For example, the question requires the intelligent agent to capture the relationships <girl, holding, hamburger>, <tray, on the left of, girl> and <food, on, tray>. The proposed Scene Graph Convolutional Network (SceneGCN) first estimates an attention distribution over the relationships (darker color indicates higher attention value) and then updates each node's hidden state by aggregating information from the relevant objects and relationships through convolutional operation.}
% \label{fig:motivation}
% \end{figure}

% (not sure if this paragraph is necessary) VQA has been studied on synthetic\cite{johnson2017clevr} and real-world images\cite{agrawal2015VQA,hudson2019GQA}. Our work focuses on real-world images based VQA since super-human performance has been achieved on synthetic datasets\cite{santoro2017simple,hudson2018MAC} and yet understanding the complex semantics in real world images remains a challenge.

However, the predominant existing solutions for VQA rely on regional features (e.g. patch-based, object-based) in image modeling \cite{yang2016stacked, lu2016hierarchical,anderson2018bottom, hudson2018MAC}. These approaches are agnostic to the visual relational clues in images. Some recent research attempted to make VQA agents more relation-aware, for example,  \cite{hudson2018MAC, santoro2017simple, teney2017Graph, wang2019neighborhood}. However, the representation of visual relationships used by these approaches are not sufficiently informative because of the following reasons: (1) Implicit models, such as Santoro \etal's Relation Network  \cite{santoro2017simple} and Hudson \etal's MAC \cite{hudson2018MAC}, are trained without explicit relational annotations and use latent or no representation for visual relationships. These models are expected to learn relational knowledge from a indirect source of supervision --- image-question-answer triples in VQA datasets. These models have limited capability of understanding rich and deep relational semantics due to the noisiness of VQA data; (2) Explicit models use explicit visual relationship representations. However, current methods either only address spatial relationships and ignore semantical relationships \cite{teney2017Graph} or use no more than labels to represent semantical relationships \cite{wang2019neighborhood, yao2018exploring}. 

Our work is inspired by advances in the area of visual relationship modeling 
\cite{lu2016visual,zhang2017visual,wan2018representation,zhang2018large}, which embed visual relationships (spatial and semantical) into vector spaces and then detect, predict relationship between objects and thereby automatically construct scene graphs on images. Comprehensively-annotated visual relationships became available with the release of large-scale visual relationship benchmarks \cite{lu2016visual, krishnavisualgenome, hudson2019GQA} and greatly facilitate visual relationship representation learning. These advances motivate us to introduce semantic-rich visual relationship modeling as a critical prior knowledge into visual relational reasoning for the VQA task. 

%Explicit visual relationship annotations\cite{lu2016visual, krishnavisualgenome, hudson2019GQA} are introduced to train models that detect, predict relationship between objects and automatically construct scene graphs on images. These advancements inspire us to introduce explicit visual relationship modeling into visual reasoning as a prior knowledge.

In this work, we try to answer two questions: (1) How to effectively represent visual relational semantics and (2) how to effectively combine and use object and relational information with a neural network. For the first question, we propose to represent images as scene graphs with each node denoting an object and each directed edge denoting a relationship between two objects. Object features in the image are extracted with an object detection network, whereas for visual relationships, we build a visual relationship encoder to yield discriminative and type-aware visual relationship embeddings constrained by both the visual context and language priors. For the second question, we propose a \textit{Scene} \textit{G}raph \textit{C}onvolutional \textit{N}etwork (\textit{SceneGCN}) to reason about the visual clues for the correct answer under the guidance of the question. The SceneGCN model mainly consists of two units: the \textit{scene graph convolution unit} unit which figures the importance of the relationships according to the question and dynamically enrich object representations by its relation-essential neighborhood; and the \textit{question guided object attention unit}  which further identifies the critical relation-aware objects to answer the question.

%a \textit{scene graph convolution unit} that updates each node's representation according to the question and the node's interaction with neighbor nodes, aggregating the information from both object properties and semantic relationships, and a \textit{question-guided object attention unit} that selects the object that is essential to answer the question.

We demonstrate the model's effectiveness, generalization, and interpretability on the challenging \textit{GQA} dataset for compositional reasoning problems as well as the popular \textit{VQA2.0} dataset for general VQA problems. It achieves state-of-the-art 54.56\% accuracy on GQA. Extensive ablation studies verifies the significance of introducing informative visual relationship priors to guide the relational reasoning process. The proposed SceneGCN model is highly interpretable. We visualize the model's progressive reasoning process on the scene graph: first localizing the important relationships according to the question and then finding essential objects referred by the question.

%The code is available from \url{https://www.acm.org/publications/proceedings-template}.   
%However, most existing benchmarks suffer from the notable language priors that the ``blind'' model can leverage the statistical biases without true reasoning skills. Besides language biase, the benchmarks commonly contains basic, non-compositional language rarely requiring reasoing capacities. To tackle these shortcomings, a new GQA dataset is recently introduced for compositional question answering based on visual reasoning.

%% file: sections/2_related_works.tex
\section{Related Work}

\subsection{Visual Question Answering}

Researchers have had different approaches to the modeling of the behavior of answering questions based on visual content. An early and typical approach fused visual and textual features for a joint representation and inferred the answer based on the fused image-question representation. Feature fusion techniques include element-wise summation/multiplication, concatenation \cite{zhou2015simple}, bilinear pooling  \cite{Tenebaum1997Separating,Fukui2016Multimodal,Kim2017Hadamard} and even more sophisticated fusion methods \cite{noh2016image}. These methods are, however, crippled by the monolithic vector representation they used for both visual and textual information which have little capability of representing fine-grained information. 

Patch-based \cite{yang2016stacked} and object-based \cite{anderson2018bottom} features are then introduced to represent information in finer granularity and thereby enrich the semantics in images that can be mined. VQA is then formed as retrieving visual information from a visual knowledge base (image) under the guidance of a textual query. These methods leveraged attention mechanisms to locate image regions that are relevant to the question.

Recently developed methods emphasize on the reasoning aspect of visual question answering --- processing and reasoning about retrieved visual information. Kim \etal~\cite{kim2018visual} address the multi-faceted nature of VQA by constructing a differentiable multi-task model in which a master module interpreted the question and queried submodules for different reasoning tasks such as counting objects, recognizing visual relationships, $etc.$. Hudson \etal~\cite{hudson2018MAC} and Chen \etal~\cite{chen2018iterative} built iterative reasoning models that mimic human's step-wise reasoning process. Santoro \etal's work is devoted specifically to discovering visual relationships via indirect supervision from visual questions and answers. Yi \etal~\cite{yi2018neural} disentangles reasoning from representation learning by first establishing structured representations of the images and questions and then performing reasoning as symbolic program execution. Narasimhan \etal~\cite{narasimhan2018out, narasimhan2018straight} studies visual question answering that requires both visual information in images and commonsense or expertise in knowledge-bases.

\subsection{Visual Relationship Modeling}

Our work is closely related to the research of visual relationship modeling, which serves as the basis of predicate prediction, visual relationship detection and automatic scene-graph construction, etc.. 
%Visual relationship modeling involves a number of tasks that require semantical visual relationship annotations, including predicate prediction, visual relationship detection and automatic scene-graph construction. 
Lu \etal~\cite{lu2016visual} introduced word embeddings as a linguistic prior, which were used for determining the plausibility of relationships.
%, \textit{e.g.}, since $\left\langle\emph{man}, \emph{ride}, \emph{horse}\right\rangle$ is plausible, it is reasonable to assume $\left\langle\emph{man}, \emph{ride}, \emph{donkey}\right\rangle$ is plausible based on the similarity between the words \emph{horse} and \emph{donkey}.
Recent approaches constantly borrow ideas and techniques from the area of knowledge embedding.
%, treating the problem of visual relationship detection as visual knowledge graph completion. 
Zhang \etal~\cite{zhang2017visual} devised the Visual Translation Embeddings framework which enforces its embeddings to fulfill $\emph{subject}+\emph{predicate}\approx\emph{object}$, a structual constraint that draws inspiration from the knowledge embedding method TransE \cite{bordes2013translating}.  
Wan \etal~\cite{wan2018representation} introduced a technique called \emph{hierarchical projection} to instantiate the conceptual embeddings (embeddings of categories rather than instances) with visual information. The instantiated embeddings are trained to fulfill $\emph{subject}_\bot+\emph{predicate}_\bot\approx\emph{object}_\bot$. This work may be interpreted as the visual counterpart of TransD \cite{ji2015knowledge}.
Zellers \etal proposed the MotifNet \cite{zellers2018neural} which mines information from motifs --- regularly appearing substructures in scene graphs.  Zhang \etal~\cite{zhang2017visual} introduces language based and embedding based supervision to cope with the long-tailed distribution of training data, which used to undermine the quality of label-based supervision. Our model builds upon the visual relationship representation inspired by \cite{zhang2018large} and studies how it can be used for VQA reasoning.

\subsection{Graph Neural Networks}

Graph Neural Networks (GNNs) extract features from topological graphs through operations in the vertice domain \cite{duvenaud2015convolutional,gilmer2017neural,yao2018exploring,velickovic2017graph} or the spectral domain \cite{kipf2016semi}. We focus on vertice-based methods since scene graphs are dynamically constructed, whereas specture-based methods operate only on static graphs.

Typical vertice-domain graph convolution \cite{duvenaud2015convolutional} on a single node takes the form of a summation over a set of states of neighbor nodes:
\begin{equation}
\textbf{h}_i^{(l+1)} = \sigma(\sum_{j \in S_i}{\textbf{W}^{(l)}\textbf{h}_j^{(l)})}
\end{equation}
where $\textbf{h}_i^{(l)}$ denote the hidden state of node $v_i$ at the $l^{th}$layer, $S_i$ is a set of nodes that are connected to node $v_i$ and $\sigma$ denote a element-wise activation function such as the $\emph{ReLU}$. This type of transformation has been shown to be very effective at accumulating and encoding features from local, structured neighborhoods, and has led to significant improvements in areas such as graph classification \cite{duvenaud2015convolutional}.

Introducing edge information into graph processing is an important problem \cite{zhou2018graph}. A number of methods represent edges with labels. Gilmer $et$ $al.$ \cite{gilmer2017neural} proposed the following equation for updating node hidden states:
\begin{equation}
\textbf{h}_i^{(l+1)} = \sigma(\sum_{j \in S_i}{f_r(\textbf{h}_i^{(l)}, \textbf{h}_j^{(l)}))}
\end{equation}
where $\textbf{h}_i^{(l)}$ denote the hidden state of node $v_i$ at the $l^{th}$layer, $S_i$ is a set of nodes that are connected to node $v_i$ and $\sigma$ denote a element-wise activation function such as the $\emph{ReLU}$. $f_r(\cdot,\cdot)$ is a function that reacts to $r$, the label of relationship between node $v_i$ and $v_j$. The type of each instance of relationship is to be determined by pre-existing information or by a relation detector. $f(\cdot,\cdot)$ is defined as a linear transformation with different transformation parameters $\textbf{W}_r^{(l)}$ for different relationships. Yao \etal~\cite{yao2018exploring} use a similar structure for image captioning.

To ameliorate this problem, they alternatively obtain $\textbf{W}_r^{(l)}$ through a linear combination of basis transformations, where the basis (a set of low-dimensional matrices) and coefficients are learned. Nevertheless, despite the reduced number of parameters, this model did not gain substantial improvements from treating each kind of relationship distinctively. We hypothesize that simply labeling the relationships is not sufficient to exploit their multitudinous, because errors in the labeling can lead to loss of useful information and introduction of noise. Additionally, parameters obtained from a basis transformation may still be too crude for generating accurate descriptions.

Another limitation of previous graph convolution models is that they use binary connectivity between nodes, that is, two nodes are either connected or uncorrelated at all. This connectivity is usually predetermined, either provided in the data or obtained with the help of priori information, such as a pretrained relation classifier \cite{yao2018exploring}. Velickovic $et$ $al.$ addressed this problem with their Graph Attention Network \cite{velickovic2017graph}, which enables contiguous  connectivity values to be dynamically determined with a multi-head attention mechanism:
\begin{equation}
\textbf{h}_i^{(l+1)} = \bigcup_{k=1}^{K}{\sigma(\sum_{j \in S_i}{\omega_{ji, k}\textbf{W}_k^{(l)}\textbf{h}_j^{(l)})}}
\end{equation}
where $k$ is the subscript of attention heads and $\bigcup$ is used to denote a method of merging attended features from different attention heads, e.g., summation, concatenation. Additionally, each attention head uses a distinctive set of parameters to encode features.

%% file: sections/4_methodology.tex
\section{Methodology}

\begin{figure*}
\centering
\includegraphics[width=0.9\textwidth]{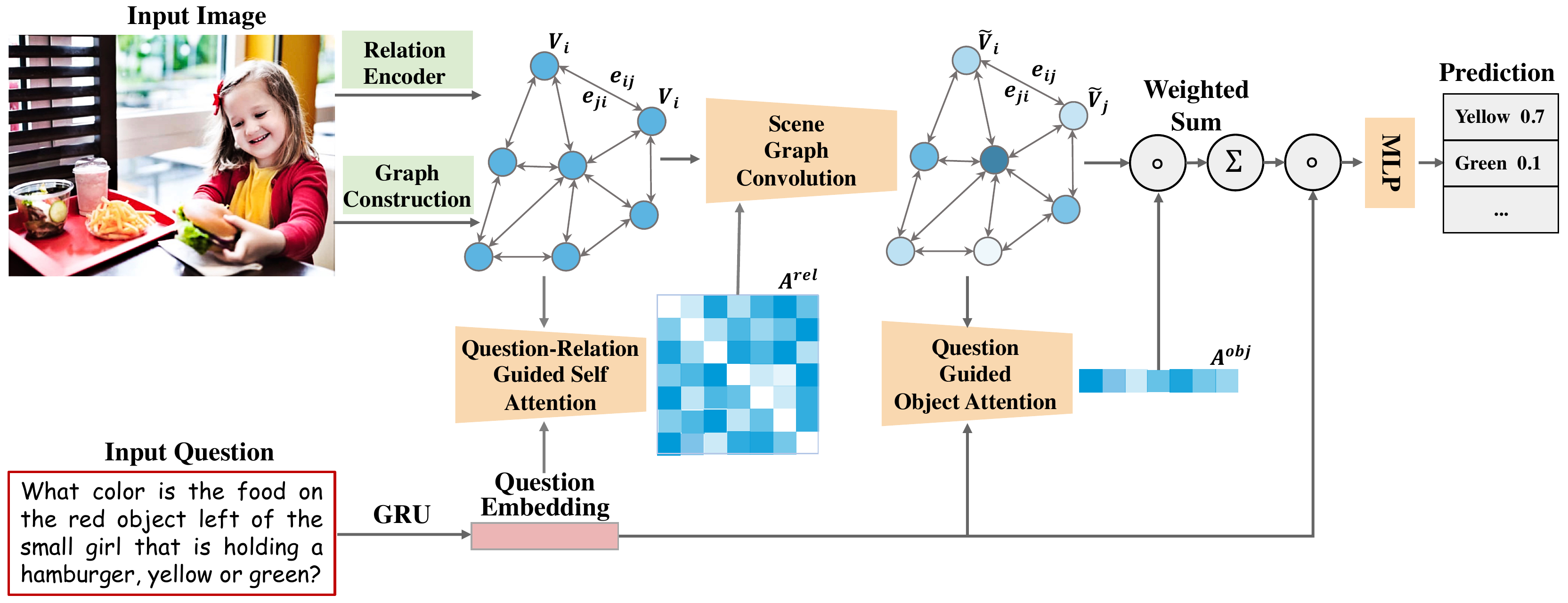} 
\caption{\textbf{Overview of the proposed SceneGCN model.} A semantic-aware scene graph is constructed from the input image first. Then the core SceneGCN module conducts a two-stage reasoning process on the scene graph over the question by enriching the node's representation (\textit{scene graph convolution}) based on the question-relevant relational prior (\textit{question-relation guided self-attention}), and aggregating the objects' information (\textit{question guided object attention}) for the image representation. The answers are predicted referring to the joint embedding of the two modalities.}
\label{fig:framework}
\end{figure*}

Our model for VQA is based on the \textit{joint embedding learning} framework which fuses the visual and textual embeddings and then uses a classifier for answer prediction. Given an image, we first construct a scene graph with embedded representation of visual objects and relationships. Then the proposed SceneGCN module conducts a two-stage reasoning process on the scene graph under the guidance of the question. The \textit{scene graph convolution unit} first dynamically updates each object's representation by aggregating the information from its neighbor nodes and semantic relationships, where the aggregation weights are learned by the question-relation guided self-attention. The \textit{question guided object attention unit} then pays different attention to the relation-aware objects and integrates the object representations based on the attention weights to form the relation-aware image representation. Such image representation is fused with the question representation via GRU and fed into MLP for answer prediction. Our framework is illustrated in Figure \ref{fig:framework}.

%Our visual question answering model is built with Anderson \etal\cite{anderson2018bottom}'s model as baseline which is under a joint-embedding framework. Questions are processed by a unidirectional, $2$-layered GRU and a question attention module\cite{yu2018beyond} to yield a single $1024$-dimensional question embedding $\textbf{q}$. Images are represented as scene graphs with each node in the graph denoting an object and each directed edge an relationship between two objects, embedded by our visual relationship encoder. The scene graph convolutional network updates each node's representation aggregating the information from both object properties and semantic relationships, in which process the connectivity between nodes is dynamically determined leveraging the relationship embeddings. Finally, a \textit{question-guided object attention} selects the object that is central to answering the question. Processed image information is fused with the question embedding and fed into a MLP to predict the final answer. The framework of our model is shown in Fig.\ref{fig:framework}

\subsection{Scene graph construction}

Each image is represented as a scene graph where the nodes, denoted as $V=\{v_i\}^{N}$, represent objects detected by a pre-trained object detector while edges, denoted as $E=\{e_{ij}\}^{N\times N}$, represent the semantic visual relationships embedded by our relationship encoder. Note that the edges are directed. We use a pre-trained object detector to detect $N$ objects in an image and describe each object as a $2,048$-dimensional vector. The relationship encoder, pre-trained on a visual relationship benchmark, i.e. \textit{GQA}  \cite{hudson2019GQA}, encodes relationships as $512$-dimensional relation embeddings, denoted as $\textbf{r}_{ij}$. We assume that certain relationship exist between any pair of objects by considering ``unknown-relationship'' as a kind of relationship. Therefore, the graph we constructed is fully-connected (except that we do not use self connections).

\subsection{Visual relationship encoder}

\begin{figure}[!b]
\centering
\includegraphics[width=0.375\textwidth]{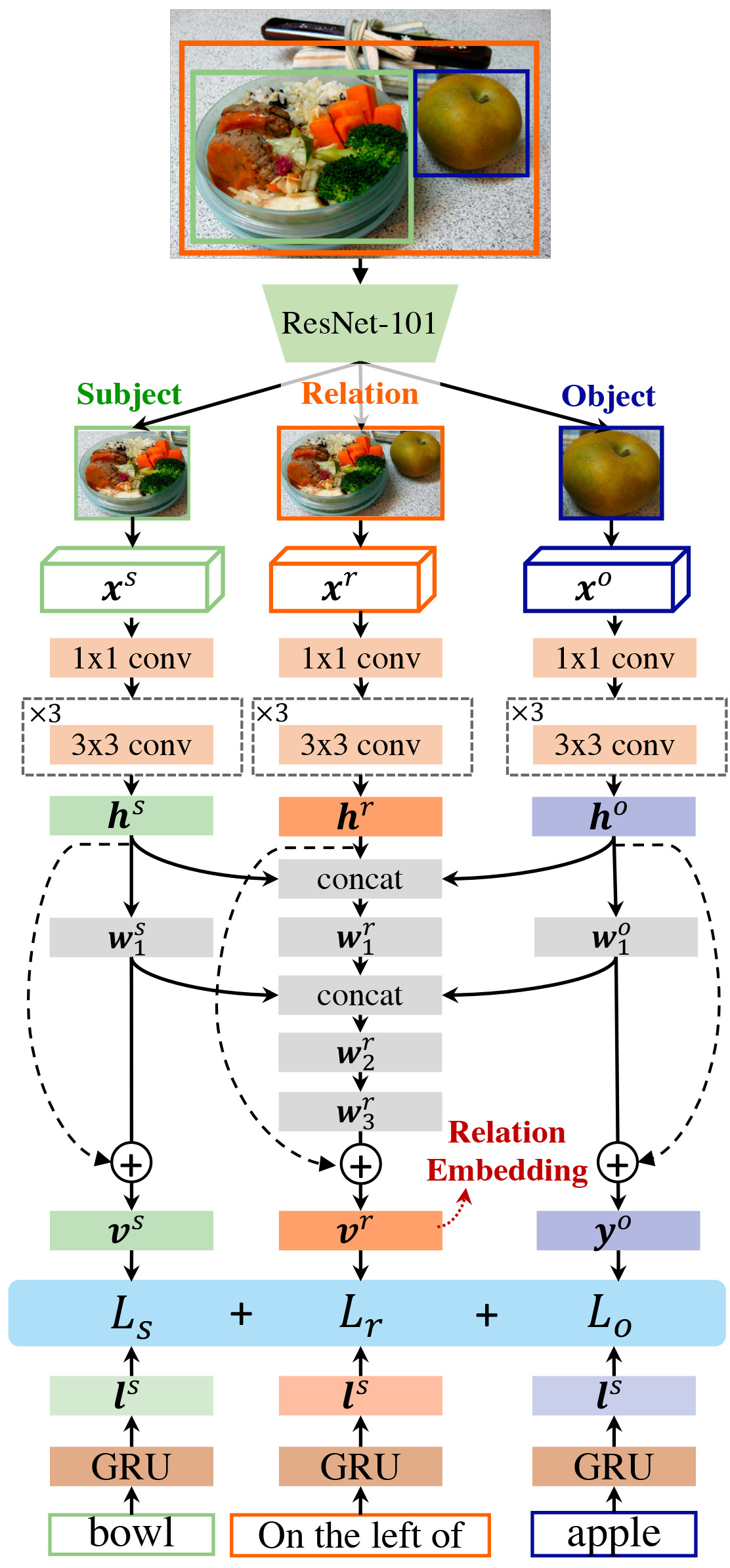}
\caption{\textbf{Illustration of the visual relationship encoder.} $L_s$, $L_o$ and $L_r$, shown in the blue box, are loss terms respectively computed form subject, object and relation embeddings. The visual module is placed above the blue box and the language module below.}
\label{fig:relationEncoder}
\end{figure}

The visual relationship encoder projects visual relationships into a deep semantic space which is aligned with their natural language annotations. This approach is different from a number of visual relationship models \cite{zhang2017visual, wan2018representation} that are trained as classifiers. The performance of those models are restricted by the long-tailed distribution of training samples among categories in scene graph datasets such as Visual Genome\cite{krishnavisualgenome} and \textit{GQA} \cite{hudson2019GQA}. Datasets are heavily trimmed to cope with this imbalance. To avoid the incomplete utilization of training data, we adopt an approach that draws inspiration from Lu \etal~\cite{lu2016visual}'s and Zhang \etal~\cite{zhang2018large}'s method which structurally aligns visual relationships with their textual annotations. The detailed structure of the visual relationship encoder is illustrated in Figure \ref{fig:relationEncoder}.

Our visual relationship encoder consists of a visual module and a language module as illustrated in Fig.\ref{fig:relationEncoder}. The visual module takes three feature maps $\textbf{x}^s$, $\textbf{x}^o$, $\textbf{x}^r$ as input and outputs three visual embedding vectors $\textbf{v}^s$, $\textbf{v}^o$, $\textbf{v}^r$ with respect to the subject, object and relationship. The language module uses a common GRU to encode the textual annotations $l^s$, $l^o$, $l^r$ to yield textual embeddings $\textbf{l}^s$, $\textbf{l}^o$, $\textbf{l}^r$ which have the same number of dimensions with the visual embeddings. Note that weights are shared between the subject-branch and the object-branch. The GRU is shared by all three branches. Images are first passed through a ResNet-101 \cite{ren2015faster} network pretrained on ImageNet \cite{deng2009imagenet} to yield $14\times14$ feature maps. Regional features are obtained through ROI-Align \cite{he2017mask} with crop-size $7\times7$. A $1\times1$ convolution layer projects the $2,048$-dimensional features down to $512$ dimensions and three consecutive $3\times3$ convolution layers compress the $7\times7$ feature maps down to a single $512$-dimensional vector for each bounding box. Fully-connected layers, denoted by $w$s in the figure, pass on to generate the final relationship embeddings. Dashed lines denote shortcut connections which are added to optimize gradient flow for the convolution layers.

The training objective is defined by 
\begin{equation}
L_\text{total} = L_s + L_r + L_o
\end{equation}
where $L_s$, $L_o$ and $L_r$ are the loss terms computed with the subject, object and relation embeddings, respectively. Next, we describe the computation of $L_s$, $L_o$ and $L_r$. The subject, object and relation subscripts are omitted since the same function is used. The loss function is designed to minimize the cosine similarity between the embeddings of positive pairs and alienate negative pairs (superscripted with ${}^{-}$), it consists of two parts:
\begin{equation}
L = L_v^{Tr} + L_l^{TrSm}
\end{equation}
where $L_v^{Tr}$ is the triplet loss:
\begin{equation}
L_v^{Tr} = \frac{1}{N_{\text{pos}} N_{\text{neg}}} \sum_{i=1}^{N_{\text{pos}}}
\sum_{j=1}^{N_{\text{neg}}}
\max[0, m-s(\textbf{v}_i, \textbf{l}_i)+s(\textbf{v}^{-}_{ij}, \textbf{l}_i)]
\end{equation}
and $L_l^{TrSm}$ is the triplet softmax loss \cite{zhang2018large}:
\begin{equation}
L_l^{TrSm} = \frac{1}{N_{\text{pos}}} \sum_{i=1}^{N_{\text{pos}}} -\log
\frac
{e^{s(\textbf{v}_i, \textbf{l}_i)}}
{
e^{s(\textbf{v}_i, \textbf{l}_i)} + 
\sum_{j=1}^{N_{\text{neg}}} e^{s(\textbf{v}_i, \textbf{l}^{-}_{ij})}
}.
\end{equation}
In these equations, we denote by $N_{\text{pos}}$ the number of positive samples in a training batch and $N_{\text{neg}}$ the number of negative samples to pair with each positive sample. The function $s(\cdot, \cdot)$ computes cosine similarity.

\subsection{Scene graph convolution}

\begin{figure}
\centering
\includegraphics[width=0.45\textwidth]{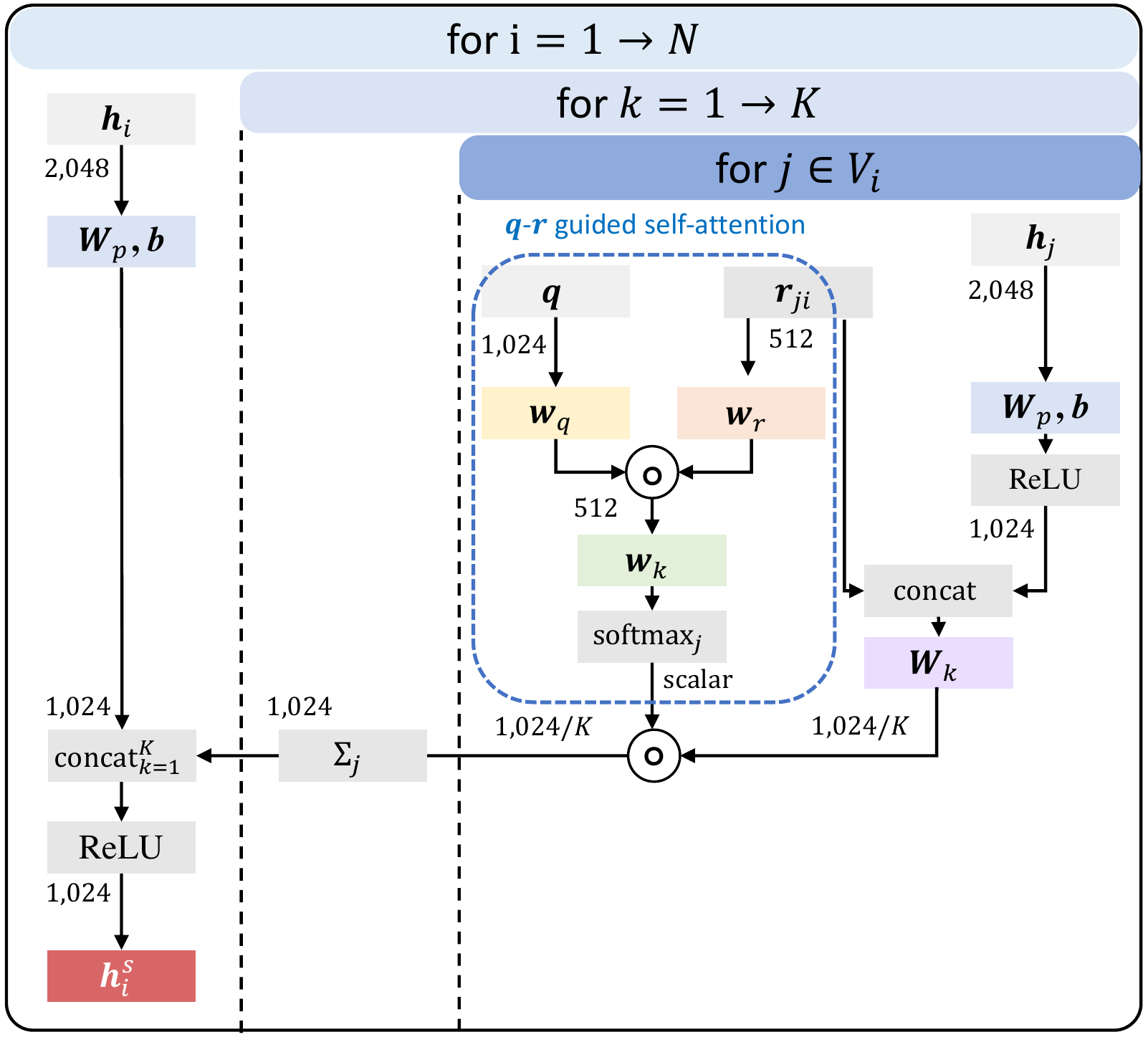}
\caption{\textbf{Illustration of the scene graph convolution operation} working in conjunction with the question-relation guided self-attention. The $\circ$ operator denotes element-wise multiplication. Different trainable parameters are annotated with different colors. The blue dashed box encircles the computation of the question-relation guided self-attention.}
\label{fig:sgconv}
\end{figure}

The scene graph convolution operation, based on graph attention networks \cite{velickovic2017graph}, updates each node's representation by aggregating information from neighbor nodes and the relationships between them. The \textit{question-relation guided self-attention} mechanism is used to infer the relevance of inter-object relationships to the question. The attention mechanism is multi-headed, allowing the operation to examine interactions among three or more objects. The complete computation process of updating node $v_i$'s hidden state through scene graph convolution is depicted in Fig.\ref{fig:sgconv}.

Denote by $\textbf{h}_i$ the hidden state of node $v_i$ and $V_i$ the set of nodes that are neighbor to $v_i$, the scene graph convolution on a single node is performed by:
\begin{enumerate}
    \item linearly projecting each object embedding (in our case from $2,048$ dimensions to $512$ dimensions):
    \begin{equation}
    \textbf{h}_{i}^{p} = \textbf{W}_p \cdot \textbf{h}_i + \textbf{b},
    \end{equation}
    where $\textbf{h}_{i}^{p}$ is the resultant projected hidden state of node $v_i$;
    \item concatenating the projected object features with relationship embeddings ($\|$ denotes concatenation):
    \begin{equation}
    \textbf{h}_{ji}^{f} = \sigma(\textbf{h}^{p}_j) ~ \| ~ \textbf{r}_{ji},
    \end{equation}
    where $\textbf{h}_{ji}^{f}$ is the concatenated vector;
    \item updating the projected object embeddings:
    \begin{equation}
    \textbf{h}^{s}_{i} = \sigma \left(\textbf{h}_{i}^{p}  +  \mathop{\|}\limits_{k=1}^{K}{\sum_{j \in V_i}{\omega_{ji, k}(\textbf{W}_{k} \cdot \textbf{h}_{ji}^{f})}}
    \right),
    \end{equation}
    where $\sigma$ denotes \textit{ReLU} activation. $\omega^{(l)}_{ji, k}$ is an attention weight --- a scalar that describes the extent to which node $v_j$ correlates to $v_i$. This attention weight is computed by the Question-relation guided self-attention. The $k$ subscript marks the subscripted parameter or attention weight as used by the $k_{\text{th}}$ attention head.
\end{enumerate}

\paragraph{Question-relation guided self-attention}
The question-relation guided self-attention (illustrated in Fig.\ref{fig:sgconv}) examines the pairwise relevance between objects in the scene graph. The attention weight from the $k^\text{th}$ attention head $\omega_{ji, k}$ measures the extent to which node $v_j$ correlates to node $v_i$. The attention weight is calculated using both the question embedding $\textbf{q}$ and the relationship embedding $\textbf{r}_{ji}$:  first, the two embedding vectors are projected and fused with
\begin{equation}
\textbf{c}_{ji} = (\textbf{w}_q \cdot \textbf{q}) \circ (\textbf{w}_r \cdot \textbf{r}_{ji})
\end{equation}
where $\circ$ denotes element-wise multiplication; then, the attention weight is obtained with
\begin{equation}
\omega_{ji, k} = \text{softmax}(\textbf{w}_k \cdot \textbf{c}_{ji} + b_k).
\label{eq:qratt}
\end{equation}

\subsection{Question guided object attention} The question guided object attention examines the updated object embeddings to find the object that is most relevant. The attention weight $\omega_{i}$ measures the importance of node $v_i$ to answering the question. It is computed by first projecting and fusing the question embedding $\textbf{q}$ with the node hidden state $\textbf{h}_{i}^{s}$
\begin{equation}
\textbf{c}^s_{i} = (\textbf{w}^s_q \cdot \textbf{q}) \circ (\textbf{w}^s_h \cdot \textbf{h}_{i}^{s})
\end{equation}
and then applying a softmax operation after a linear layer
\begin{equation}
\omega_{i} = \text{softmax}(\textbf{w}^s \cdot \textbf{c}^s_{i} + b^s).
\label{eq:qatt}
\end{equation}

In the end, all visual information in the graph nodes are aggregated by
\begin{equation}
\hat{\textbf{v}} = \sum_{i=1}^N \omega_{i} \cdot \textbf{h}_{i}^{s}.
\end{equation}
to yield $\hat{\textbf{v}}$, a vector that is expected to contain the relevant visual information for answering the question.

\subsection{Answer Prediction}
We employ a two-layer MLP to predict the $\textit{scores}$ for candidate answers. Formally,
\begin{equation}
    \text{scores} = \sigma(\text{MLP}(\hat{\textbf{v}} \circ \textbf{w}_a\textbf{q}))
\end{equation}
The MLP has $1,024$ and $N_a$ units in its first and second layer and uses a dropout probability of $0.5$, where $N_a$ is the number of all possible answers. We denote the predicted score for the $i^{th}$  answer by  $s_i$ and the benchmark label by $\hat{s}_{i}$. %, detailed in \ref{sec:exp}.
The binary cross-entropy loss we use is formally defined by:
\begin{equation}
    L = \sum_{j=1}^{B} \sum_{i=1}^{N_a} ((1-s_{i}^{(j)})\text{log}(1-\hat{s}_{i}^{(j)}) - s_{i}^{(j)}\text{log}(\hat{s}_{i}^{(j)}))
\end{equation}
where $B$ denotes the batch-size and $j$ subscripts question-answers entries in training batches. Although \textit{GQA} annotates each question with only one answer, it is empirically found that binary loss works better than softmax loss.

\subsection{Implementation details}

\paragraph{Object Features} We directly use the object features provided by Anderson \etal\cite{anderson2018bottom} for \textit{VQA 2.0} and for \textit{GQA}\cite{hudson2019GQA} we use the features that is provided in the dataset.

\paragraph{Visual Relationship Encoder} The visual relationship encoder is trained $5$ epochs on \textit{GQA} \cite{hudson2019GQA}  using an Adam \cite{kingma2014adam} optimizer with learning rate $0.0001$, learning rate decay $0.8$ and weight decay $0.0001$. $300$-dimensional $Glove$ \cite{pennington2014glove} word embeddings are passed through a unidirectional, $1$-layered GRU to obtain embeddings of the subject's name, object's name and predicate. $m$ in the loss function is set $0.2$. $N_{\text{pos}}$ and $N_{\text{neg}}$ are set $256$ and $128$, respectively. It is reported in \cite{zhang2017visual} that cosine similarity values need to be manually scaled before computing the triplet softmax loss in order to prevent gradient vanishing, the adapted model does not benefit from this technique. All $311$ predicates are used to train the visual relationship encoder albeit each predicate is limited to provide a maximum of $10,000$ training samples.

\paragraph{Visual Question Answering} We train the visual question answering model on \textit{GQA} balanced train questions with an Adamax \cite{kingma2014adam} solver for $20$ epochs. Batch size and learning rate are set $256$ and $0.001$, respectively. Gradients with norm exceeding $0.25$ are clipped. The object-based features we use are the ones provided in \textit{GQA} and we use a maximum of $N=36$ objects for each image. The number of all possible answers $N_a$ for \textit{GQA} is set to $1,298$ by filtering out answers that appear less than $20$ times as they constitute a tiny portion of all the questions. For \textit{VQA 2.0}, we filter out answers that appear less than $9$ times following \cite{anderson2018bottom}.

%% file: sections/5_experiments.tex
\section{Experiments}

\begin{table*}
\begin{center}
\caption{Ablation study of SceneGCN on \textit{GQA} validation split.}
\label{table:ablation_val}
 \begin{tabular}{ c | c | c c | c c | c } 
 \toprule
 Metric & Baseline & + width & + \emph{q} att & + SceneGCN & + 2 att & implicit \\
 \midrule
 Accuracy & 59.07 & 60.83 & 61.03 & \textbf{62.45} & 62.42 & 60.50 \\ 
 \rowcolor{lightblue}
 Open & 45.47 & 46.69 & 46.52 & 47.88 & \textbf{47.93} & 46.25 \\
 \rowcolor{lightblue}
 Binary & 73.58 & 75.92 & 76.51 & \textbf{77.99} & 77.89 & 75.70 \\
%  Query & & & & & & & \\
%  Compare & & & & & & & \\
%  Choose & & & & & &   & \\
%  Logical & & & & & & & \\
%  Verify & & & & & & & \\
 Global & 64.64 & 65.01 & 65.46 & 65.70 & \textbf{66.20} & 64.89 \\
 Object & 80.15 & 80.90 & 81.59 & \textbf{82.14} & 81.99 & 81.86 \\
 Attribute & 62.18 & 65.95 & 66.07 & \textbf{68.56} & 68.39 & 65.21 \\
 \rowcolor{lightyellow}
 Relation & 52.01 & 52.73 & 52.92 & 54.04 & \textbf{54.25} & 52.45 \\
 Category & 53.63 & 55.47 & 55.15 & \textbf{55.55} & 54.54 & 54.28 \\
 \rowcolor{lightblue}
 Distribution & 6.01 & 5.31 & 4.95 & 5.58 & 6.35 & \textbf{4.43} \\
 Grounding & 80.39 & 89.84 & 90.89 & 92.32 & \textbf{93.19} & 87.26 \\
 \rowcolor{lightblue}
 Validity & 94.77 & \textbf{95.01} & 94.97 & 94.92 & 94.97 & 94.95 \\
 \rowcolor{lightblue}
 Plausibility & 90.88 & 91.31 & 91.24 & 91.23 & \textbf{91.40} & 91.10 \\
 Consistency & 84.14 & 84.27 & 84.75 & 87.14 & \textbf{88.17} & 83.43 \\ 
 \bottomrule
\end{tabular}
\end{center}
\end{table*}

\begin{table*}
\begin{center}
\caption{Comparison of SceneGCN with State-of-the-art on \textit{GQA} test split.}
\label{table:sota_gqa}
 \begin{tabular}{ c | c c c c c c c }
 \toprule
 Model & Acc. & Open & Binary & Dist. & Validity & Plaus. & Consist. \\
 \midrule
%  Global Prior\cite{hudson2019GQA} & 28.93 & 16.52 & 42.99 & 130.86 & 89.02 & 75.34 & 51.78 \\
 Local Prior \cite{hudson2019GQA} & 31.31 & 16.99 & 47.53 & 21.56 & 84.44 & 84.42 & 51.34 \\
  \rowcolor{lightblue}
%  LSTM\cite{hudson2019GQA} & 41.07 & 22.69 & 61.90 & 17.93 & 96.39 & 87.30 & 68.68 \\
  \rowcolor{lightblue}
%  CNN\cite{hudson2019GQA} & 17.82 & 1.74 & 36.05 & 19.99 & 35.78 & 34.84 & 62.40 \\
  \rowcolor{lightblue}
 CNN+LSTM \cite{hudson2019GQA} & 46.55 & 31.80 & 63.26 & 7.46 & 96.02 & 84.25 & 74.57 \\\
 BottomUp \cite{anderson2018bottom} & 49.74 & 34.83 & 66.64 & 5.98 & \textbf{96.18} & \textbf{84.57} & 78.71 \\
 \rowcolor{lightblue}
 MAC \cite{hudson2018MAC} & 54.06 & 38.91 & \textbf{71.23} & \textbf{5.34} & 96.16 & 84.48 & 81.59 \\
 \rowcolor{lightyellow}
 SceneGCN (ours) & \textbf{54.56} & \textbf{40.63} & 70.33 & 6.43 & 95.90 & 84.23 & \textbf{83.49} \\
 \midrule
 Human \cite{hudson2019GQA} & 89.3 & 87.4 & 91.2 & - & 98.9 &97.2 & 98.4 \\
 \bottomrule
\end{tabular}
\end{center}
\end{table*}

\begin{table*}
  \centering
  \begin{threeparttable}
  \caption{Comparison of SceneGCN with State-of-the-art on \textit{VQA 2.0} test split.}
  \label{table:sota_vqa2}
  \setlength{\tabcolsep}{0.8mm}{
    \begin{tabular}{lcccccccccccccc}
    \toprule
    \multirow{2}{*}{Model}&
    \multicolumn{4}{c}{test-dev}&\multicolumn{4}{c}{test-standard}\cr
    \cmidrule(lr){2-5} \cmidrule(lr){6-9}
    &  Overall & Other & Number  & Yes/No & Overall & Other & Number  & Yes/No \cr
    \midrule
   % \textbf{State of the arts:}&&&&&&&\cr
    Prior \cite{goyal2017making} & - & -&-&-& 25.98 & 01.17 & 00.36 & 61.20\\
    \rowcolor{lightblue}
    % Language only \cite{goyal2017making} & - & -&-&-& 44.26 & 27.37 & 31.55 & 67.01\\
    LSTM+CNN \cite{goyal2017making}& - & -&-&-& 54.22 & 41.83 & 35.18 & 73.46 \\
    \rowcolor{lightblue}
    MCB  \cite{goyal2017making} & - & -&-&-& 62.27 & 53.36 & 38.28 & 78.82\\
    % DCN \cite{inprovedFusionForVQA} & 66.60 & 56.72 & \textbf{46.60} & \textbf{83.50} & 67.00 & 56.90 & \textbf{46.93} & \textbf{83.89}\\
    BottomUp \cite{anderson2018bottom} & 65.32 & 56.05 & 44.21 & 81.82 & 65.67 & 56.26 & 43.90 & 82.20\\
    \rowcolor{lightblue}
    LV-NUS\cite{ilievski2017simple} & - & - & - & - & 66.77 & \textbf{58.30} & 46.29 & 81.89 \\
    \rowcolor{lightyellow}
    SceneGCN (ours) & 66.81 & 57.77 & 46.85 & 82.72 & \textbf{67.14} & 57.89 & \textbf{46.61} & \textbf{83.16} \\
    \bottomrule
    \end{tabular}}
    \end{threeparttable}
\end{table*}

We present an ablation study to examine the effectiveness of the proposed method, a comparison with state-of-the-art methods and a qualitative evaluation. 

\subsection{Datasets and Metrics}

\textit{GQA} and \textit{VQA 2.0} are used to evaluate our visual question answering model. The ablation study is conducted solely on \textit{GQA} as the dataset provides a comprehensive set of evaluation metrics including one devoted to measuring agents' capability of understanding visual relationships. We compare with state-of-the-art methods on both datasets.

\paragraph{GQA} \cite{hudson2019GQA} is a new dataset for VQA over real-world images with fixes of the flaws of existing benchmarks --- strong language priors, use of basic, non-compositional language and ambiguous refering. $52\%$ of the questions in \textit{GQA} are annotated to require an understanding of visual relationships. The effectiveness of our model can be directly reflected on the \textit{relation} questions in \textit{GQA}. Apart from the standard accuracy metric measured on a number of different categories of questions, the dataset also provides the \textit{Consistency} metric which measures responses on semantically equivalent or entailing questions, the \textit{Validity} and \textit{Plausibility} metrics which test whether answers are within a reasonable range, the \textit{Distribution} metric which measures the difference between the predicted answer distribution and the real, and the \textit{Grouding} metric which checks whether the model attends to regions within the image that are relevant to the question.

\paragraph{VQA 2.0} \cite{agrawal2015VQA} is a popular visual question answering benchmark with measures taken against guessing answers basing solely on language priors. \textit{VQA 2.0} reports accuracy on three categories of questions: \textit{yes/no}, \textit{number} and \textit{other}. Note that for \textit{VQA 2.0} the groundtruth score of an answer is determined by a vote among $10$ human annotators:
\begin{equation}
    \text{score}_\text{gt}(a_i) = \min \big\{ 1, \frac{n_i}{3} \big\}
\end{equation}
where $n_i$ is the times that an answer $a_i$ is voted by different annotators.

% The ablation study is conducted on \textit{GQA}, a new dataset for VQA over real-world images with fixes of the flaws of existing benchmarks --- strong language priors, use of basic, non-compositional language and ambiguous refering. A suite of new metrics are released to complement the standard accuracy measure. $52\%$ of the questions in \textit{GQA} are annotated to require an understanding of visual relationships. The effectiveness of our model can be directly reflected in the \textit{relation} metric measured by \textit{GQA}. The reset of the experiments are conducted on both \textit{GQA} and \textit{VQA 2.0}.

% All experiments are conducted on a server with 4 Tesla V100 GPUs. For the \textit{grounding} metric, we report the scores measured using attention weights computed by the \textit{question guided object attention} module. In the state-of-the-art comparison, the models are trained on the combination of the \textit{train} and \textit{val} split.

\subsection{Ablation study}

In this experiment, improvements are progressively added to the model. The models in this experiments are trained on the balanced train split of \textit{GQA} and evaluated on the complete validation split. We use the
BottomUp \cite{anderson2018bottom} as baseline (with the minor adaptation that \textit{ReLU} is used instead of gated tanh activation) and add improvements one after another. Numeric results are shown in Table.\ref{table:ablation_val}.

\paragraph{Effectiveness of improvements on the baseline} The \textit{+ width} model uses increased embedding size. The size of the question embeddings and visual features are increased from $512$ to $1024$. The \textit{+ \emph{q} att} model improves the question embedding module. We pass the question through a 2-layer GRU (instead of 1-layer) and a question attention module \cite{yu2018beyond} to obtain attentive question features. These improvements account for $1.96$\% of overall accuracy improvement. The \textit{relation} metric is improved by $0.91$\%.

\paragraph{Effectiveness of the scene graph convolution} Two models with our scene graph convolution are tested. The \textit{+ SceneGCN} model uses a single attention head in the scene graph convolution and the \textit{+ 2 att} model uses 2 attention heads. The SceneGCN module further improves the accuracy and \textit{relation} score by $1.41$\% and $1.33$\% respectively. Besides, most of the metrics are improved when SceneGCN is employed. The SceneGCN module with two attention heads benefits a number of different metrics compared with the one with a single attention head, despite not achieving the highest overall accuracy. Using more attention heads improves performance on \textit{open}-ended questions, \textit{global} and \textit{relation}al understanding, \textit{grounding} and \textit{consistency}, proving that the SceneGCN module is capable of improving reasoning abilities.

\paragraph{Comparison with implicit relational reasoning} To examine the degree to which the SceceGCN improves relational understanding, we compare with the \textit{implicit} model which uses the concatenation of object feature vectors as relation embeddings and is otherwise the same as \textit{+ SceneGCN}. This practive is similar with \cite{santoro2017simple}. The results show that the \textit{implicit} relational reasoning model does not bring apparent improvement to the model, we hypothesize the main reason to be the non-deterministic nature of visual relationships --- it is insufficient to determine visual relationships given only the properties of the subject and the object.

\subsection{State-of-the-art comparison}

The comparison between our method with state-of-the-art methods is shown in Table.\ref{table:sota_gqa} and Table.\ref{table:sota_vqa2}. Results on both datasets show that the SceneGCN model improves significantly over the baseline \textit{BottomUp}  \cite{anderson2018bottom}, by $4.82$\%  and $1.47$\% of overall accuracy, respectively. The performance on \textit{open}-ended questions is notably improved by $5.80$ on \textit{GQA}. Our model outperforms the state-of-the-art model \textit{MAC} \cite{hudson2018MAC} on overall accuracy, the ability to answer \textit{open}-ended questions and \textit{consistency}, proving the advantage of prior visual relationship learning over implicit methods. Also, the model trained and evaluated on \textit{VQA 2.0} gained less of an improvement compared to the one on \textit{GQA}, we hypothesize this is because a much smaller portion of questions in \textit{VQA 2.0} tests relational understanding.

\subsection{Interpretability}

\begin{figure*}
\centering
\includegraphics[width=0.675\textwidth]{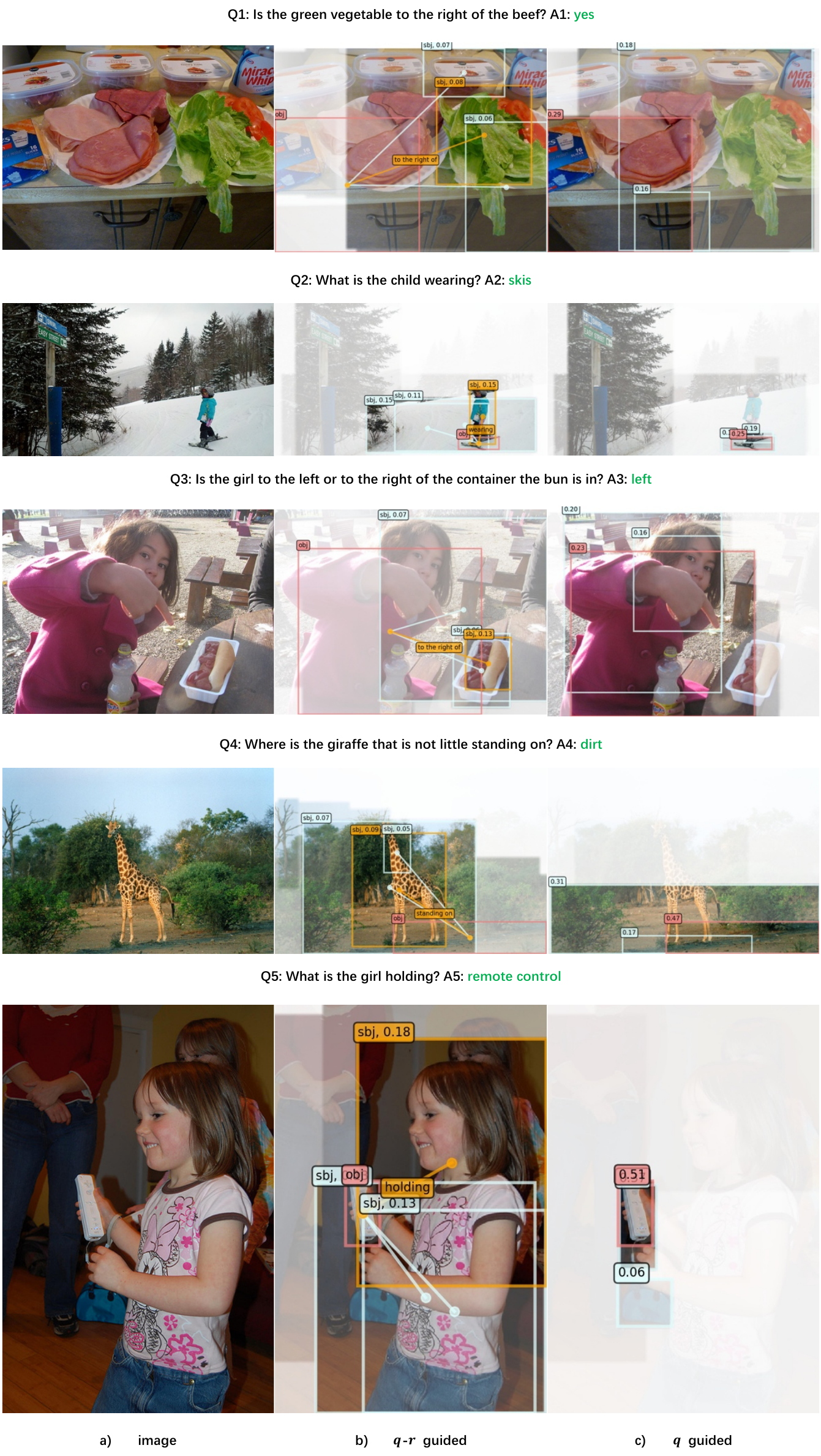} 
\caption{\textbf{Qualitative Results.} For each question, the first column shows the original image. The second column depicts the top three relationships of a \textit{central object} according to the attention weights learned by the question-relation guided self-attention ($\omega_{ji,k}$ as in Eq.\ref{eq:qratt}). The relationship with the highest attention weight is marked with orange, whereas light blue is used to mark the $2^\text{nd}$ and $3^\text{rd}$ most relevant relationships. The \textit{central object} is the one with the highest weight from the question guided object attention ($\omega_{i}$ as in Eq.\ref{eq:qatt}). We also show the relation label predicted by the visual relationship encoder in the second column. The third column visualizes the distribution of the the question guided object attention, where the \textit{central object} is marked with red.}
\label{fig:vis_1}
\end{figure*}

\begin{figure*}
\centering
\includegraphics[width=0.675\textwidth]{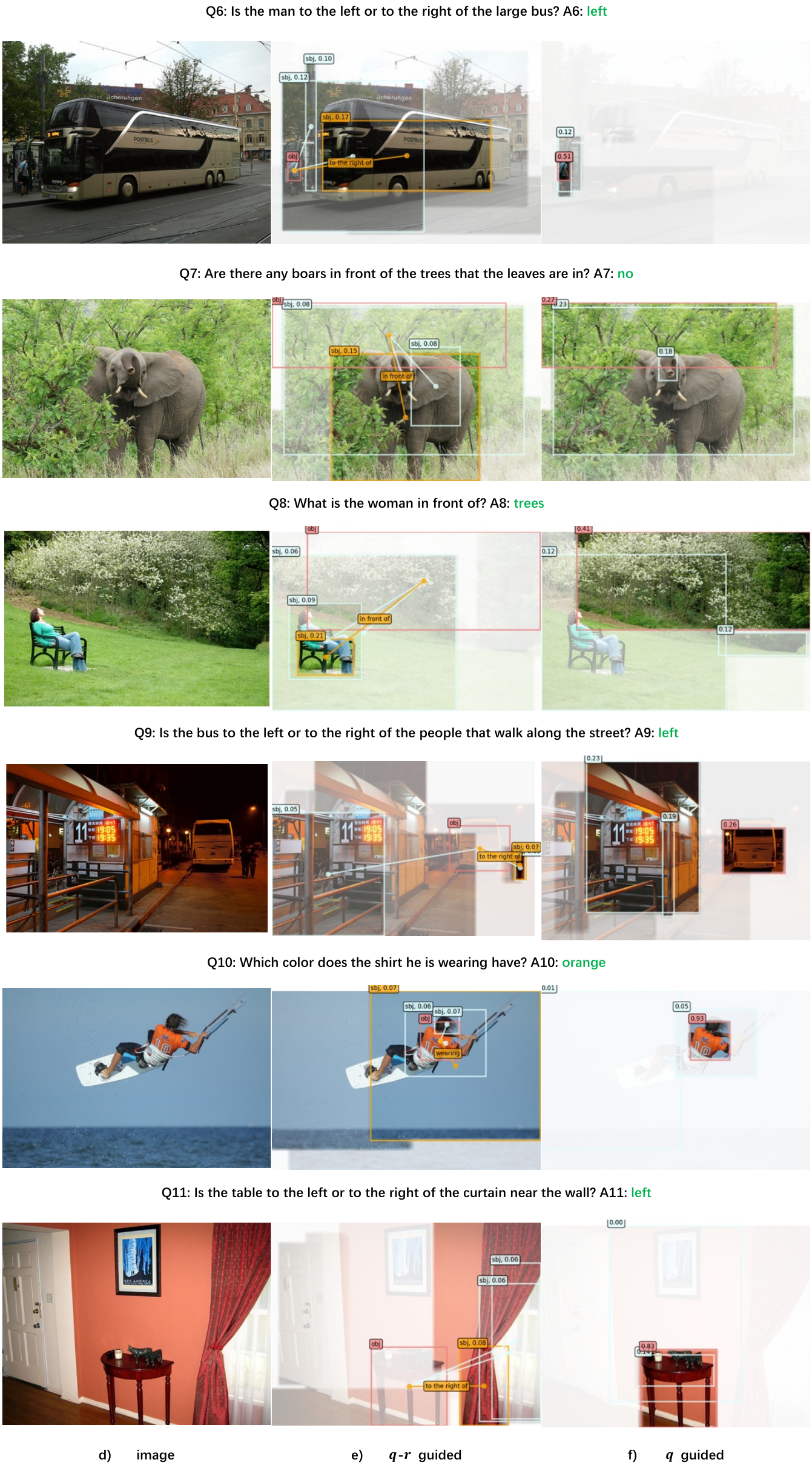} 
\caption{\textbf{Qualitative Results.} For each question, the first column shows the original image. The second column depicts the top three relationships of a \textit{central object} according to the attention weights learned by the question-relation guided self-attention ($\omega_{ji,k}$ as in Eq.\ref{eq:qratt}). The relationship with the highest attention weight is marked with orange, whereas light blue is used to mark the $2^\text{nd}$ and $3^\text{rd}$ most relevant relationships. The \textit{central object} is the one with the highest weight from the question guided object attention ($\omega_{i}$ as in Eq.\ref{eq:qatt}). We also show the relation label predicted by the visual relationship encoder in the second column. The third column visualizes the distribution of the the question guided object attention, where the \textit{central object} is arked with red.}
\label{fig:vis_2}
\end{figure*}

\begin{figure*}
\centering
\includegraphics[width=0.675\textwidth]{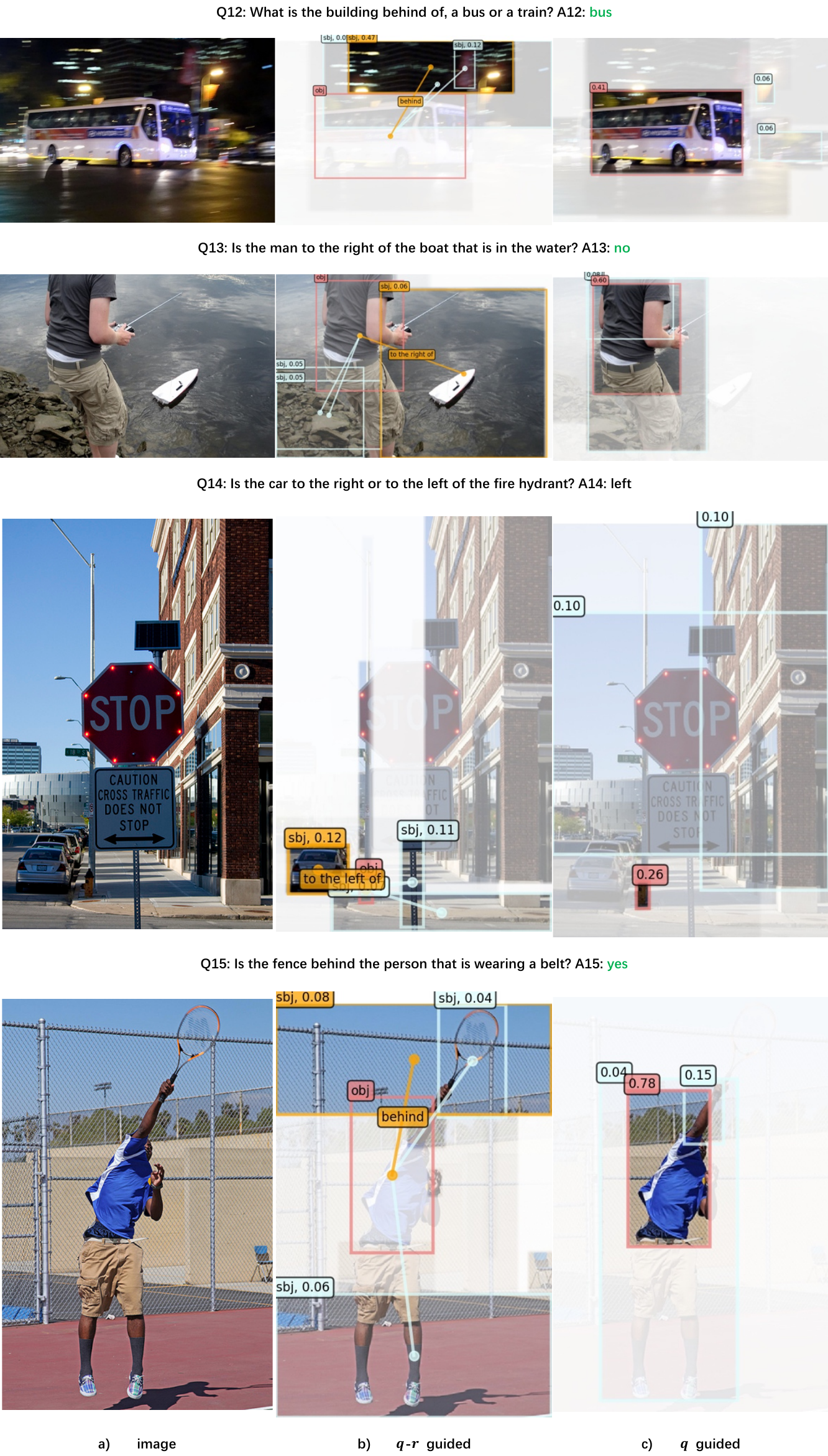} 
\caption{\textbf{Qualitative Results.} For each question, the first column shows the original image. The second column depicts the top three relationships of a \textit{central object} according to the attention weights learned by the question-relation guided self-attention ($\omega_{ji,k}$ as in Eq.\ref{eq:qratt}). The relationship with the highest attention weight is marked with orange, whereas light blue is used to mark the $2^\text{nd}$ and $3^\text{rd}$ most relevant relationships. The \textit{central object} is the one with the highest weight from the question guided object attention ($\omega_{i}$ as in Eq.\ref{eq:qatt}). We also show the relation label predicted by the visual relationship encoder in the second column. The third column visualizes the distribution of the the question guided object attention, where the \textit{central object} is marked with red.}
\label{fig:vis_3}
\end{figure*}

To obtain insight into the reasoning process and how SceneGCN leverages the relational information, we visualize the attention distributions produced by the model (with a single attention head in the scene graph convolution) during the progressive computation and show examples in Figure \ref{fig:vis_1}, \ref{fig:vis_2} and \ref{fig:vis_3}.

We observe that the question-relation guided self-attention is capable of capturing the critical relationships for the correct answers. It performs well on various kinds of relationships, including spatial, interactive and comparative ones. Moreover, the attention maps of the question guided object attention demonstrate the model's ability to focus on the most relevant objects based on the relation-aware object representations. We take the first example in Figure \ref{fig:vis_1} to see how the model explicitly reasons about the relation-level and object-level clues: first identifying the most relevant relationship \textit{to the right of} from all the candidate relationships for the subject \textit{green vegetable}, then locating on the target object \textit{beef} serving as strong evidence for predicting the answer \textit{yes}. In the second step, notice how the model attends to the \textit{beef} instead of other objects. This owes to the ability of the model to conduct progressive reasoning and aggregating beneficial relational clues from the prior step, which guides it to only focus on the most critical objects referring to the question.

\subsection{Experimental Details}

All experiments are conducted on a server with 4 Tesla V100 GPUs. For the \textit{grounding} metric, we report the scores measured using attention weights computed by the \textit{question guided object attention} module. 

Following previous work, we train our model on the training set and report the results on the validation set in our ablation study. We train our model on both the training and validation set and report the results returned by the evaluation servers for the state-of-the-art comparison.

%% file: sections/6_conclusion.tex
\section{Conclusion}

In this paper, a novel solution to the problem of relational reasoning in visual question answering was proposed by utilizing prior visual relationship learning. The new model is named as Scene Graph Convolutional Network, by which input images are represented by structured scene graphs using a pretrained object detector and a pretrained visual relationship encoder that embeds both objects and relationships. The proposed scene graph convolution operation updates each node's hidden states using information from both objects and relationships. We demonstrated the effectiveness and transparency of the model through quantitative and qualitative studies, and achieved state-of-the-art results on both the \textit{GQA} and \textit{VQA 2.0} datasets. Morover, interpretable attention-based visualization shows strong evidence that our model has good capability in relationship identifying and progressive reasoning.